\documentclass[%
 reprint, superscriptaddress,
 amsmath,amssymb,
 aps,
prl,
floatfix,
]{revtex4-2}

\usepackage{graphicx}
\usepackage{dcolumn}
\usepackage{bm}
\usepackage{hyperref}
\usepackage[mathlines]{lineno}
\usepackage{graphicx}
\usepackage{multirow}
\usepackage{array}
\usepackage[T1]{fontenc}
\usepackage{xcolor}
\usepackage{subcaption}
\usepackage{hyperref}

\begin{document}

\preprint{APS/123-QED}

\title{From cell intercalation to flow, the importance of T1 transitions}

\author{Harish P. Jain}
 \affiliation{Njord Centre, Physics Department, University of Oslo}
 
\author{Axel Voigt}
\affiliation{Institute of Scientific Computing, TU Dresden, 01069 Dresden, Germany}
\affiliation{Center for Systems Biology Dresden, Pfotenhauerstr. 108, 01307 Dresden, Germany}
\affiliation{Cluster of Excellence, Physics of Life, TU Dresden, Arnoldstr. 18, 01307 Dresden, Germany}

\author{Luiza Angheluta}
\affiliation{Njord Centre, Physics Department, University of Oslo}

\begin{abstract}
Within the context of epithelial monolayers, T1 transitions, also known as cell-intercalations, are topological rearrangements of cells that contribute to fluidity of the epithelial monolayers. We use a multi-phase field model to show that the ensemble-averaged flow profile of a T1 transition exhibits a saddle point structure, where large velocities are localised near cells undergoing T1 transitions, contributing to vortical flow. This tissue fluidisation corresponds to the dispersion of cells relative to each other. While the temporal evolution of the mean pair-separation distance between initially neighbouring cells depends on specific model details, the mean pair-separation distance increases linearly with the number of T1 transitions, in a way that is robust to model parameters.   
\end{abstract}

\maketitle

\textit{Introduction.- } 
Spontaneous flows within epithelial monolayers play a pivotal role in many biological processes including tissue development, wound healing, angiogenesis and invasion of cancer cells \cite{Ladoux2017, bruguesForcesDrivingEpithelial2014, friedlClassifyingCollectiveCancer2012}. In confluency, cells may migrate collectively while maintaining contact with their neighboring cells, thus preserving cell-cell junctions \cite{Ladoux2017}. However, cells may also migrate relative to each other through structural rearrangements mediated by local events of remodelling of cell-cell junctions. Three types of such events are identified~\cite{weaireEffectStrainTopology1992, etournayInterplayCellDynamics2015, guiraoUnifiedQuantitativeCharacterization2015}. A T1 transition is an event whereby two neighbouring cells move apart while two of their neighbours move towards each other and make contact. A T2 transition occurs during extrusion/apoptosis events where cells are eliminated from the monolayer, and a T3 transition is an event associated with cell division. Notably, T1 events preserve the total cell count within the monolayer, distinguishing them from T2 and T3 transitions. 

T1 transitions have been observed in both epithelial \cite{irvineCellIntercalationDrosophila1994} and mesenchymal \cite{kellerMechanismsConvergenceExtension2000} tissues, and play an active role in development of an embryo from early gastrulation to late-stage organogenesis \cite{walck-shannonCellIntercalationTop2014}, as well as during cancer metastasis \cite{oswaldJammingTransitionsCancer2017a}. Empirical evidence shows that Myosin II contributes to the build-up of tension at cell-cell junctions, thereby controlling T1 transitions in epithelial tissues \cite{bertetMyosindependentJunctionRemodelling2004, takeichiDynamicContactsRearranging2014a}. However, the interplay between mechanical and biochemical signaling in these events remains a topic of debate \cite{rauziCellIntercalationSimple2020}. There have been several studies that are aimed at characterising the mechanical influence of T1 transitions in tissues \cite{rauziNatureAnisotropyCortical2008, guiraoUnifiedQuantitativeCharacterization2015, blanchardTissueTectonicsMorphogenetic2009, aigouyCellFlowReorients2010, etournayInterplayCellDynamics2015, curranMyosinIIControls2017,kimEmbryonicTissuesActive2021a} and in simulations \cite{blanchardTissueTectonicsMorphogenetic2009, jainRobustStatisticalProperties2023a, biDensityindependentRigidityTransition2015, bartonActiveVertexModel2017}.

Tissue flow has been explored mostly using the coarse-grained approaches from active nematics or active polar matter or, more generally, active p-atic liquid crystals, whereby spontaneous flows are influenced by their topological defects \cite{alertActiveTurbulence2022, doostmohammadiActiveNematics2018, armengol-colladoEpitheliaAreMultiscale2023}. Several experimental studies report different discrete orientational order (nematic, polar or in general p-atic) of cell tissues, corresponding to different lowest-energy topological defects~\cite{duclosTopologicalDefectsConfined2017,Saw2017,armengol-colladoEpitheliaAreMultiscale2023}. This system dependency may be attributed to different types of cell lines, but may also signal that the system is far from a hydrodynamic limit with a well-defined discrete symmetry, thus making the identification of relevant group symmetries and corresponding topological defects a matter of debate. On the other hand, it is becoming more evident that T1 transitions are important sources of tissue flow \cite{Lecuit2015, jainRobustStatisticalProperties2023a}. Qualitative differences between flows occurring at hydrodynamic scales and those induced by structural re-arrangements of cells have been identified by comparing hydrodynamic (continuum) simulations with cell-resolved (discrete) simulations \cite{doostmohammadiActiveNematics2018,wenzelDefectsActiveNematics2021}. How the flows originating at the discrete cell level influence the tissue flow at hydrodynamic scales is far less explored and understood.  

In this paper, we aim to bridge this gap to better understand the generic flow patterns due to cell neighbor rearrangements. Within a two-dimensional multi-phase field model, resolving the dynamics of each individual cell, we  quantify the average flow profile of a T1 transition. We demonstrate that T1 transitions are short-lived and localised events of cells with high speeds generating a $4$-fold vorticity with alternating chirality. Considering the number of cells and their size to remain constant, by using a Lagrangian approach, we quantify how the cell pair dispersion is mediated solely by T1 transitions. We predict a robust scaling of the mean pair separation with the number of T1 transitions which hints at a generic mechanism of relative dispersion by structural rearrangements. The details of the model are in the Supplementary Material (SM) which includes the Refs.~\cite{nonomuraStudyMulticellularSystems2012, Camley2014, Lober2015, marthCollectiveMigrationHydrodynamic2016, muellerEmergenceActiveNematic2019, wenzelMultiphaseFieldModels2021, shenDiffuseInterfaceModel2023, happelCoordinatedMotionEpithelial2024, Vey2007, Witkowski2015, praetoriusCollectiveCellBehaviour2018}.

\begin{figure*}[]
    \centering
    \begin{subfigure}[b]{0.8\textwidth}
        \centering
        \includegraphics[width=\textwidth]{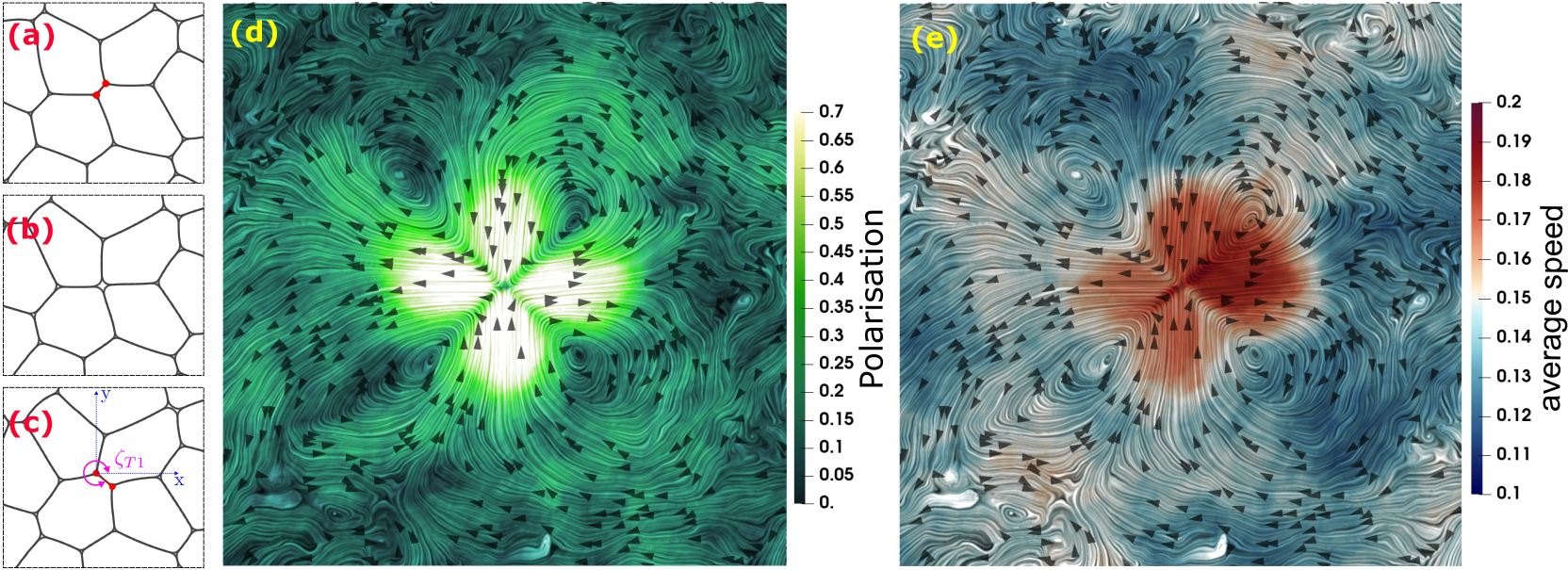}
    \end{subfigure}
    \caption{Configuration of $4$ cells (a) before, (b) during, and (c) after a T1 transition. The 3-way vertices involved in the T1 are marked by red dots. Angle $\zeta$ is the orientation of the newly-formed junction  after a T1 transition measured counterclockwise from the positive $x$ semi-axis. Streamlines of the average flow profile at the end of a T1 transition (d) and (e). The colormap in (d) represents the magnitude of the polarisation, $p$. $p=1$ at a location if for all T1 transitions in the ensemble, the velocity points in the same direction at that location, and $p=0$ if velocity is randomly oriented. The colormap in (e) represents the average speed. The flow is visualized using LIC (line integral convolution), and the arrows indicate the direction. The evolution of these fields before and after the T1 transition are shown in the Supplementary Movies 1-4, see SM \cite{SM}.}
    \label{fig:T1profiles}
    
\end{figure*}


\begin{figure*}
    \centering
    \includegraphics[width=.95\textwidth]{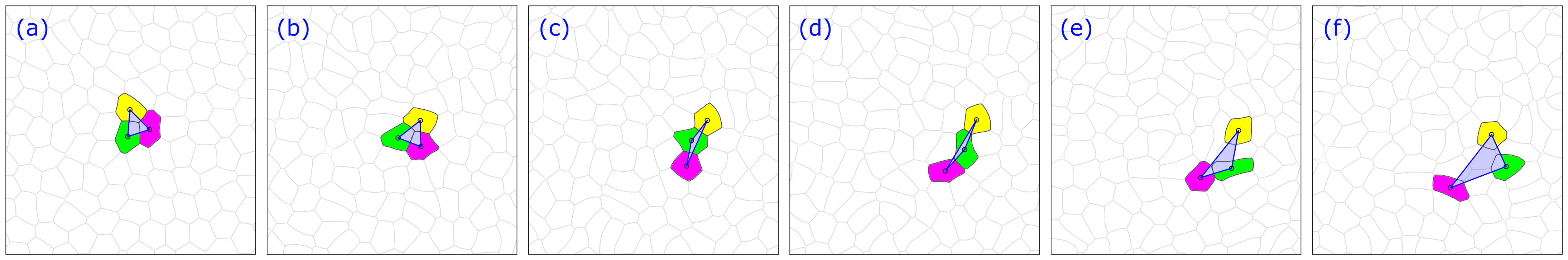}
    \caption{Temporal snapshots of the evolution of a cell triplet. T1 transitions break the contact between cells in the triplet, and increase the area spanned by the triplet. }
    \label{fig:triangle_evolution}
\end{figure*}

\paragraph{T1 transitions.- } A T1 transition re-configures the junctions between neighboring cells as illustrated in Fig.~(\hyperref[fig:T1profiles]{1a, 1b, 1c}). The two 3-way vertex represented by red dots move towards each other shrinking the corresponding junction (Fig.~\hyperref[fig:T1profiles]{1a}). Within multi-phase field models, a T1 transition is a spatial-temporal event with finite duration \cite{jainRobustStatisticalProperties2023a}, that starts when the junction vanishes, i.e. two 3-way vertices merge creating a transient extracellular gap (approximated as a four-way vertex), as shown in Fig.~\hyperref[fig:T1profiles]{1b}. The T1 transition concludes when a new pair of 3-way vertices nucleate forming a new junction. We define the orientation, $\zeta$, of the T1 transition by the angle of the newly formed junction with respect to the x-axis (Fig.~\hyperref[fig:T1profiles]{1a}). We use this angle $\zeta$ to reorient the different T1 transitions and the corresponding velocity profile in order to compute ensemble-averaged properties as detailed in the SM~\cite{SM}. The corresponding flow field is reoriented such that the junction created after the T1 transition aligns with the horizontal axis, see Fig.~(\hyperref[fig:T1profiles]{1d, 1e}). In these figures, we illustrate the average flow profile at the end of a T1 transition. This shows the saddle point structure with two orthogonal directions of attraction (along the vertical axis) and repulsion (along the horizontal axis), respectively. In turn, this is a source of quadruple vortices with alternating circulations. The colormap in Fig.~(\hyperref[fig:T1profiles]{1d}) corresponds to the magnitude of polarization, $p(\mathbf x,t)$, which varies spatially between $0$ and $1$. The polarization is $1$ when on average cell velocities point in the same direction at a given position $\mathbf x$. Conversely, $p(\mathbf x,t) = 0$, if there is no preferred direction at $\mathbf x$. Notice that the non-zero polarization is localised around the T1 transition in such a way that we can discern the ensemble-averaged shape of the $4$ cells involved in the T1 transition. The polarisation field informs on the likelihood of the observed streamlines, where a higher polarisation indicates that the streamlines shown are more likely to occur. The colormap in Fig.~(\hyperref[fig:T1profiles]{1e}) corresponds to the ensemble-average of the magnitude of the velocity field around a T1 transition. Higher magnitudes are localised around the center, where the cells involved in the T1 transition are present, and decreases in the rest of the domain \cite{Palmieri2015, jainRobustStatisticalProperties2023a}. The  movies $1-4$ from the SM~\cite{SM} show the evolution of flow profiles before and after a T1 transition. The temporal behavior is consistent with the cell speed profile before and after a T1 transition~\cite{jainRobustStatisticalProperties2023a}, whereby the typical cell speed increases during a T1 transition, and decays back to the average speed of all cells after the event. This property along with the flow profile suggests that T1 transitions are localised burst of high speeds, leading to vortical flow \cite{jainRobustStatisticalProperties2023a}. 
Cells that are caged among their neighbours, undergo small shape fluctuations and experience resistance to their motion due to intercellular interaction. However, these small fluctuations may build up and can bring two 3-way vertices closer together. This leads to accumulation of energy near those vertices. Eventually, this can induce a T1 transition releasing energy, and thereby helping uncage the cells. This is akin to the buildup prior to a slip event in stick-slip dynamics~\cite{sethna2001crackling}. Provided, there is no preferred orientation of the T1 transition, the cells would all move away from their initial neighbours as elucidated in the next section.

\textit{Relative Dispersion.- } The vortical flow induced by T1 transitions is reflected in the dispersion of cells relative to each other, and thus in the mixing of cells. We can quantify the flow properties indirectly through the relative dispersion of cells. We use the center of mass denoted by $\mathbf{r}_i$ for the $i$-th cell to track cell migration, see SM~\cite{SM} for details. To quantify single particle dispersion, we use the mean-squared displacement of the cells given as
\begin{equation}
    \sigma^2(t) = \frac{1}{N}\sum\limits_{i=1}^N |\mathbf r_i(t+t_0)-\mathbf r_i(t_0)|^2,
\end{equation}
which measures how far a cell has migrated from its initial location at $t_0$ in a time-lag $t$. $N$ is the total number of cells. Likewise, to quantify relative dispersion, we define the mean pair-separation distance in a time-lag $t$ as 
\begin{equation}
    \Pi(t) = \frac{1}{N_{P}(t_0)}\sum\limits_{j=1}^{N_{P}(t_0)} |\mathbf r_{j, 1}(t)-\mathbf r_{j, 2}(t)|,
\end{equation}
where $N_{P}(t_0)$ is the number of pairs of cells that were neighbours at an initial time $t_0$. $\mathbf r_{j, 1}$ and $\mathbf r_{j, 2}$ are the positions of the two cells corresponding to the $j^{th}$ neighbour pair at $t_0$. 
To better quantify the dynamics of the 3-way vertices, we also consider the dispersion of triplets. A triplet consists of $3$ cells where each cell is a neighbour to the other two cells. Let $\mathbf r_{k, 1}$, $\mathbf r_{k, 2}$, and $\mathbf r_{k, 3}$ be the positions of the $3$ cells corresponding to the $k^{th}$ triplet. The corresponding lengths of the edges in the triplet triangle are $\bar{a} = |\mathbf r_{k, 1}-\mathbf r_{k, 2}|$, $\bar{b} = |\mathbf r_{k, 2}-\mathbf r_{k, 3}|$, and $\bar{c} = |\mathbf r_{k, 3}-\mathbf r_{k, 1}|$.  Given the perimeter of the triangle, $\bar{s} = \bar{a}+\bar{b}+\bar{c}$, the triangle area of the $k^{th}$ triplet is  $A_{k}= \bar{s}(\bar{s}-\bar{a})(\bar{s}-\bar{b})(\bar{s}-\bar{c})$. Thus, we can also define the mean triplet-separation area as 
\begin{equation}
    \Lambda(t) = \frac{1}{N_{T}(t_0)}\sum\limits_{k=1}^{N_{T}(t_0)} A_k(t+t_0),
\end{equation}
where $N_{T}(t_0)$ is the number of triplets of cells at time $t_0$. $\Lambda(t)$ is the average area of all triangles made by triplet of cells that were mutually adjacent at time $t_0$. Snapshots from the evolution of one such triplet triangle are shown in Fig~\ref{fig:triangle_evolution}. Geometrically, the dispersion of cells is associated with deformations, translations and rotations of the triplet triangle.

The initial area of the triplet triangles is non-zero and is related to the cell size. Notice that the cells remain mutually adjacent for a transient time until the 3-way vertex shared between them is destroyed in a T1 transition by merging with another 3-way vertex. From hereon, the cells are likely to further separate from each other due to subsequent T1 transitions.  The triplet area could also collapse to $0$ when the $3$ cells become collinear (see Fig.~\hyperref[fig:triangle_evolution]{2c-d}).

\begin{figure}[]
    \centering
    \begin{subfigure}[b]{0.49\textwidth}
        \centering
        \includegraphics[width=\textwidth]{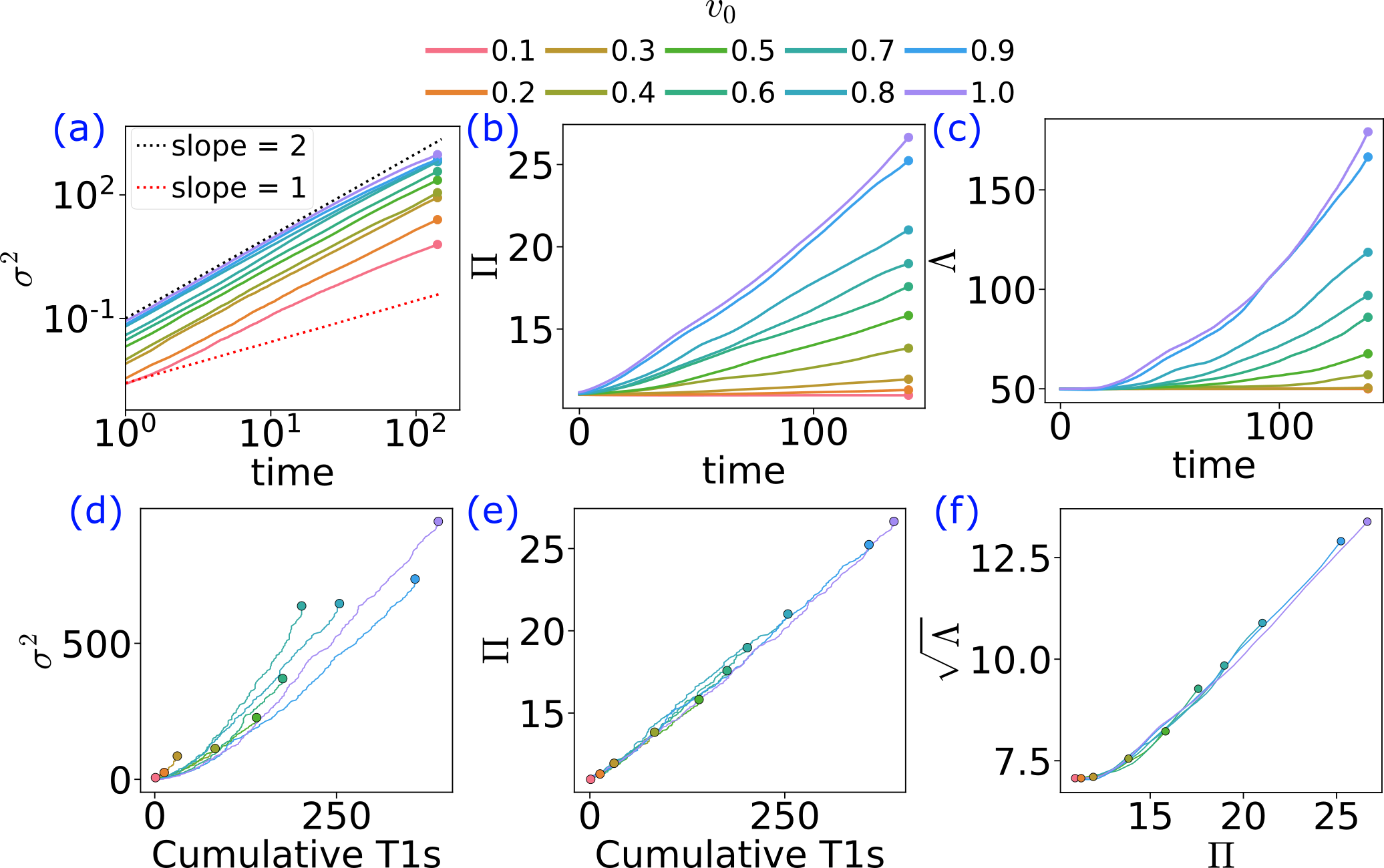}
    \end{subfigure}
        \caption{Evolution of $\sigma^2$ (a), $\Pi$ (b) and $\Lambda$ (c) plotted against time. $\sigma^2$ (d) and $\Pi$ (e) plotted against the cumulative number of T1 transitions within the tissue.  $\sqrt{\Lambda}$ plotted against $\Pi$ (f). Different colours correspond to different cell activities defined by the self-propulsion speed $v_0$ (see SM~\cite{SM}), as per the legend at the top. }
    \label{fig:dispersion properties}
    \end{figure}

The mean dispersion of single cells ($\sigma^2$), cell pairs ($\Pi$) and triplets ($\Lambda$) as functions of time-lag $t$ and for different activities $v_0$, see SM \cite{SM}, are shown in Fig.~\hyperref[fig:dispersion properties]{3a}, \hyperref[fig:dispersion properties]{3b},  and \hyperref[fig:dispersion properties]{3c}. Within the considered computational time, we see that cells are super-diffusive with a scaling exponent that lies between $1$ (normal diffusion limit of Brownian motion) and $2$ (ballistic limit of self-propulsion). For all three quantities, the net dispersion increases with cell activity which is consistent with the observation that the fluctuations in the tissue flow increase with activity \cite{alertActiveTurbulence2022}. At low activities, we notice that there is almost no dispersion of cell pairs and triplets. While, the pair-separation $\Pi$ increases with time-lag, the triplet area $\Lambda(t)$ remains constant for a initial transient period. A single T1 transition would destroy the 3-way vertex shared by the cells in the triplet, but $2$ pairs out of $3$ initial pairs in the triplet remain. The triplet area starts to diverge when at least one of the cells breaks contact with the other $2$ cells in the triplet, which requires a minimum of $2$ T1 transitions. Thus, in this initial period, the area of the triplet triangle $A_k(t)$ fluctuates around $A_k(t_0)$. As the cells undergo further T1 transitions and move apart, the area spanned by the triplet $\Lambda$ increases monotonically with the time-lag.

Fig.~\hyperref[fig:dispersion properties]{3d} and ~\hyperref[fig:dispersion properties]{3e} show the net dispersion of single cells ($\sigma^2(t)$) and cell pairs ($\Pi(t)$) plotted against the cumulative number of T1 transitions within the tissue in time-lag $t$. Interestingly, the curves for $\Pi$ for different $v_0$ collapse into a linear master curve. A similar collapse is also present for the triplet area, which shows that these two quantities are directly related to the number of T1 transitions. However, this is not the case for $\sigma^2$, because cells can also migrate in local flocks, which affects $\sigma^2$, but does not change $\Pi$ and $\Lambda$ since the cell-cell connectivity is preserved \cite{Ladoux2017}. In the absence of T1 transitions, cells are topologically caged within their neighbours, and this makes it difficult to generate flow fluctuations (when T2 and T3 transitions are also absent). It takes approximately $250$ T1 transitions across the tissue of $100$ cells, or equivalently $10$ T1 transitions per cell, for the mean-pair separation to double its value. Two neighbouring cells have to undergo around $10$ T1 transitions each, such that their cell centers could be two cell apart. When $\sqrt{\Lambda}$ is plotted against $\Pi$, initially $\sqrt{\Lambda}$ is constant, and later it varies linearly with $\Pi$ (see Fig.~\hyperref[fig:dispersion properties]{3f}). 

\begin{figure}[]
    \centering
    \begin{subfigure}[b]{0.49\textwidth}
        \centering
        \includegraphics[width=\textwidth]{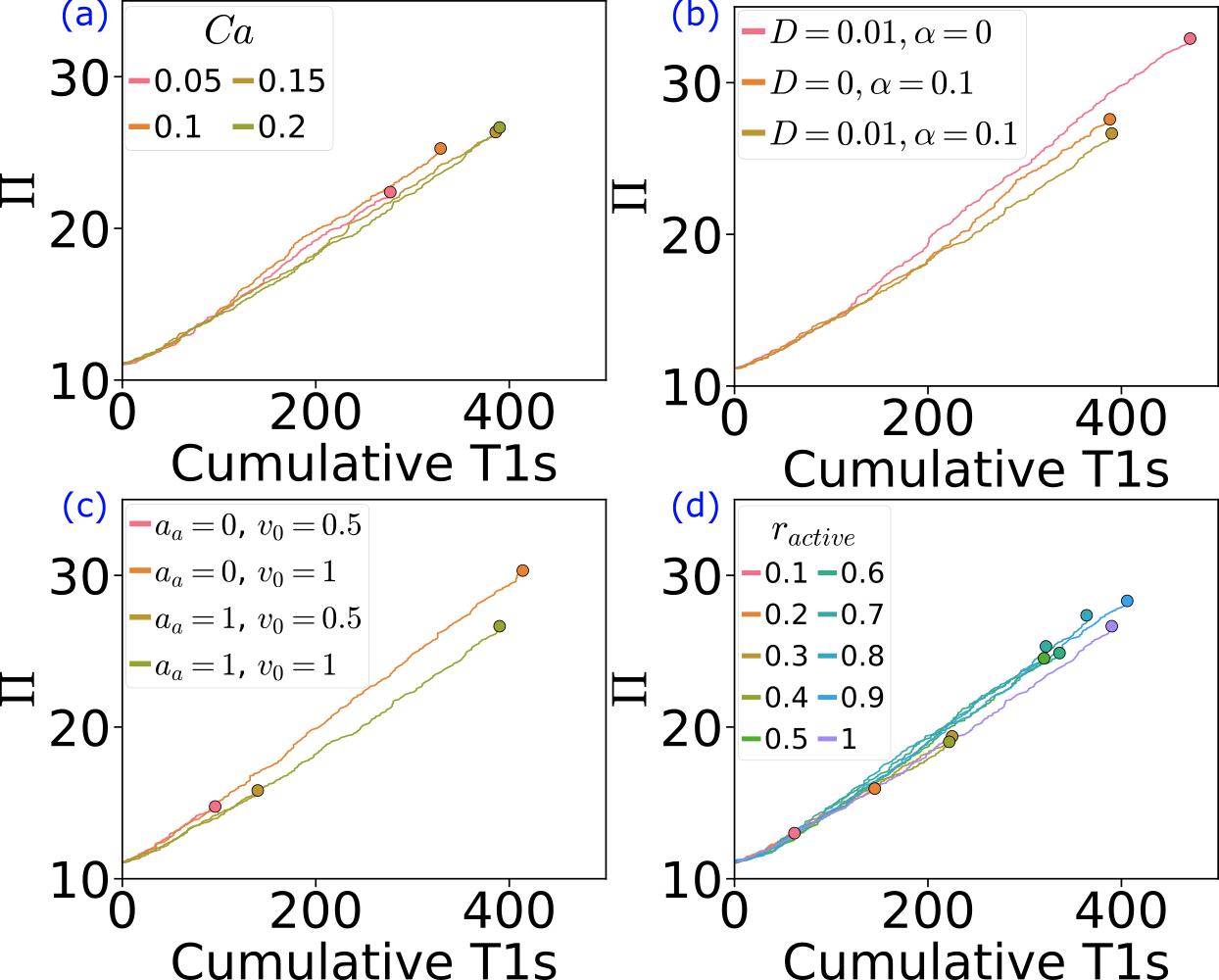}
    \end{subfigure}
    \caption{Data collapse of the relative dispersion curves as function of the cumulative T1 transitions for various model parameters and setups: (a) cell deformability ($Ca$) is varied for $v_0 =1$. (b) rotational noise $D$ and shape alignment $\alpha$ are varied for $v_0 =1$. (c) repulsion vs repulsion and adhesion cell-cell interaction. (d) mixtures of active and passive cells with a fraction $r_{active} $ of active cells with $v_0=1$.    
    } \label{fig:cumulative_T1_Pi}
    \end{figure}

To further explore the robustness of this data collapse, we also vary other model parameters i.e. cell deformability $Ca$, rotational noise $D$, shape alignment $\alpha$, adhesion parameter $a_a$, and activity ratio $r_{active}$. For high values of $Ca$, the cells are more deformable. The rotational diffusion coefficient $D$, and the rate $\alpha$ at which cells tend to align their preferred direction of motion with the direction of their elongation control the dynamics of the self-propulsion mechanism. For $a_a=0$, the interactions between cells are purely repulsive, while $a_a>0$ corresponds to additional attraction interaction between cells. $r_{active}$ is the fraction of cells which have nonzero activity. See SM for details \cite{SM}. The mean-pair separation distance as a function of cumulative number of T1 transitions for these scenarios is plotted in Fig.~(\ref{fig:cumulative_T1_Pi}). In all these scenarios, we see similar scaling behaviour suggesting that T1 transitions affect relative dispersion in similar ways irrespective of model parameters. 

\textit{Discussion.- }
In summary, we have studied how T1 transitions contribute to the fluidization of an epithelial monolayer. We have shown that T1 transitions, as topological events of neighbour exchanges, are sources of transient saddle-point flows generating 4-fold vortices. The average polarization is high around the principle axes of the saddle point, which corresponds to underlying cells losing and gaining neighbours. Due to these flows, T1 transitions promote cell mixing. While the temporal behavior of the cell dispersion properties (single, pair or triplets) depends on the model parameters, we found an underlying robust scaling law of the \emph{relative} dispersion (pairs and triplets) as function of the number of T1 transitions. The data collapse of the different dispersion curves onto a linear master curve suggests that the rate of relative dispersion is directly proportional of the occurrence rate of T1 transitions regardless of the underlying nucleation mechanism. This unveils a deeper connection between topological rearrangements and mixing in densely packed systems, in general. As $\Pi(t)$ and the cumulative number of T1 transitions are measurable within epithelial tissues and other computational models, our results are accessible for experimental validation.

\begin{acknowledgments}
L.A. and H.P.J. acknowledge funding by the European Union’s Horizon $2020$ research and innovation programme under the Marie Skłodowska-Curie grant agreement No 945371. A.V. acknowledges funding by the German Research Foundation (DFG) within grant FOR3013. We are grateful for discussions with Amin Doostmohammadi, Richard D.J.G. Ho, Per Arne Rikvold and Dag K. Dysthe. 
\end{acknowledgments}

\bibliography{mainT1flow}

\end{document}


\preprint{APS/123-QED}



\title{Supplementary material}

\author{Harish P. Jain}
 \affiliation{Njord Centre, Physics Department, University of Oslo}
 
\author{Axel Voigt}
\affiliation{Institute of Scientific Computing, TU Dresden, 01069 Dresden, Germany}
\affiliation{Center for Systems Biology Dresden, Pfotenhauerstr. 108, 01307 Dresden, Germany}
\affiliation{Cluster of Excellence, Physics of Life, TU Dresden, Arnoldstr. 18, 01307 Dresden, Germany}

\author{Luiza Angheluta}
\affiliation{Njord Centre, Physics Department, University of Oslo}

\maketitle

\section{Model, Parameters and Computation}
We consider a multi-phase field model, see Ref.~\cite{nonomuraStudyMulticellularSystems2012,Camley2014,Lober2015,Palmieri2015,marthCollectiveMigrationHydrodynamic2016,muellerEmergenceActiveNematic2019} for various realizations and Ref.~\cite{wenzelMultiphaseFieldModels2021} for a review and comparison. The specific formulation builds on the one used in Ref.~\cite{jainRobustStatisticalProperties2023a} and is used to simulate the dynamics of a confluent monolayer of $N=100$ cells. Each cell is described by a phase field variable $\phi_i$ with $i=1, 2, \ldots, N$, which has the bulk values $\phi_i \approx 1$ inside the cell, $\phi_i \approx -1$ outside it, with a diffuse interface of width ${\cal{O}}(\epsilon)$ between them representing the cell boundary. The phase field $\phi_i$ follows the dynamics 
\begin{equation}\label{eq: Phase field evolution}
    \partial_t \phi_i + \mathbf{v}_i \cdot \nabla \phi_i = \Delta \frac{\delta  \mathcal{F}}{\delta \phi_i},
\end{equation}
which ensures conservation of cell area. Diffusive relaxation mechanisms are considered by the total free energy functional $\mathcal{F} = \mathcal{F}_{CH} + \mathcal{F}_{REP} + \mathcal{F}_{ADH} $. It contains the Cahn-Hilliard energy
\begin{equation}\label{eq: Cahn Hilliard Energy}
    \mathcal{F}_{CH} = \frac{1}{Ca}\sum_{i=1}^N \int_{\Omega} \left(\frac{\epsilon}{2}||\nabla \phi_i||^2 + \frac{1}{4\epsilon}(\phi_i^2-1)^2\right) d\mathbf{x},
\end{equation}
with the capillary number $Ca$ to tune the cell deformability, and two interaction energies, where 
    \begin{equation}\label{eq: Repulsion Energy}
        \mathcal{F}_{REP} = \frac{a_r}{2In}\sum_{i=1}^N \int_{\Omega}\sum_{j\neq i} (\phi_i+1)^2(\phi_j+1)^2 d\mathbf{x}
    \end{equation}
is the repulsion energy to prevent overlap of cell interior, and 
    \begin{equation}\label{eq: Adhesion Energy}
        \mathcal{F}_{ADH} = \frac{a_a}{2In}\sum_{i=1}^N \int_{\Omega}\sum_{j\neq i} (\phi_i^2-1)^2(\phi_j^2-1)^2 d\mathbf{x}
    \end{equation}
is the adhesion energy that promotes overlap of cell boundaries. $In$ denotes the interaction strength, and $a_r$ and $a_a$ are parameters to tune contribution of repulsion and adhesion energy. The adhesive part differs from \cite{jainRobustStatisticalProperties2023a} by rewriting the interaction using the equilibrium condition $\frac{\epsilon}{2} \|\nabla \phi_i\|^2 \approx \frac{1}{4 \epsilon}(\phi_i^2 -1)^2$, see \cite{guSimulatingVesicleSubstrate2014, marthCollectiveMigrationHydrodynamic2016, jainImpactContactInhibition2022, shenDiffuseInterfaceModel2023, happelCoordinatedMotionEpithelial2024}.

Cell activity is introduced through the advection term. We consider $\mathbf{v}_i(\mathbf{x}, t) = v_0 B(\phi_i) \mathbf{e}_i(t)$, where $v_0$ is controlling the magnitude of activity, $B(\phi_i) = \frac{\phi_i+1}{2}$ and $\mathbf{e}_i= [\cos \theta_i (t), \sin  \theta_i(t) ]$ is its direction. The migration orientation $\theta_i(t)$ evolves diffusively with drift that aligns to the principal axis of cell's elongation as 
\begin{equation}\label{eq: evolution of theta}
    d\theta_i = \sqrt{2 D_r }dW_i(t) + \alpha(\beta_i(t) - \theta_i(t))dt,
\end{equation}
with rotational diffusivity $D_r$ and a Wiener process $W_i$. $\beta_i(t)$ is the orientation of the cell elongation and is defined as

 \begin{equation}
 \beta_i(t) = \left\{ 
 \begin{array}{rl}
 \arg(\boldsymbol{\eta}_i^{+}(t)) &: \;\; \mathbf{e}_i(t)\cdot \mathbf{\eta}_i^{+}(t) > 0 \\ -\arg(\boldsymbol{\eta}_i^{+}(t)) &: \;\; \mathbf{e}_i(t)\cdot \mathbf{\eta}_i^{+}(t) < 0 .
 \end{array}\right.
 \end{equation}
The cell elongation is identified by the principal eigenvector $\boldsymbol{\eta}^{+}_i$ of the shape deformation tensor as described in Ref.~\cite{muellerEmergenceActiveNematic2019}. The parameter $\alpha$ controls the time scale of this alignment. Setting $v_0 = 0$ for some cells allows to model mixture of active self-propelled cells and passive cells. The activity ratio, $r_{active}$, is the fraction of cells for which $v_0>0$.   

Eqs. \eqref{eq: Phase field evolution} are considered in a square domain $\Omega = [0,L] \times [0,L]$ (where $L=100$) with periodic boundary conditions for the time $[0, T]$. 

The problem is solved numerically using finite elements within the toolbox AMDiS \cite{Vey2007, Witkowski2015} and the parallelization concept introduced in \cite{praetoriusCollectiveCellBehaviour2018}, which considers each cell on a different core and accounts for short-range interaction between cells to reduce the communication. This essentially allows scaling with the number of cells. The system of partial differential equations is discretized in time using a semi-implicit approach. Timestep size and mesh resolution are related to ensure stability. For further details and numerical tests, see \cite{wenzelMultiphaseFieldModels2021,jainImpactContactInhibition2022,jainRobustStatisticalProperties2023a}.

The parameters are as per Table~\ref{tab: parameters} unless specified otherwise.

    \begin{table}[htb!]
    \centering
    \begin{tabular}{|c|c|}
        \hline
        Parameter & Value \\
        \hline
        $L$ & 100 \\
        $T$ & 150 \\
        $\epsilon$ & 0.15\\
        $v_0$ & 0.5\\
        $Ca$ & 0.2\\
        $In$& 0.1\\
        $a_a$& 1\\
        $a_r$& 1\\
        $D_r$& 0.01\\
        $\alpha$ & 0.1\\
        $r_{active}$ & 1 \\
        $\tau_n$ & 0.005 \\
        \hline
    \end{tabular}
    \caption{Default values of the model parameters. $\tau_n$ denotes the timestep size.}
    \label{tab: parameters}
    \end{table}

\section{Periodic boundary conditions and dispersion}

\begin{figure}
    \centering
    \includegraphics[width=.8\textwidth]{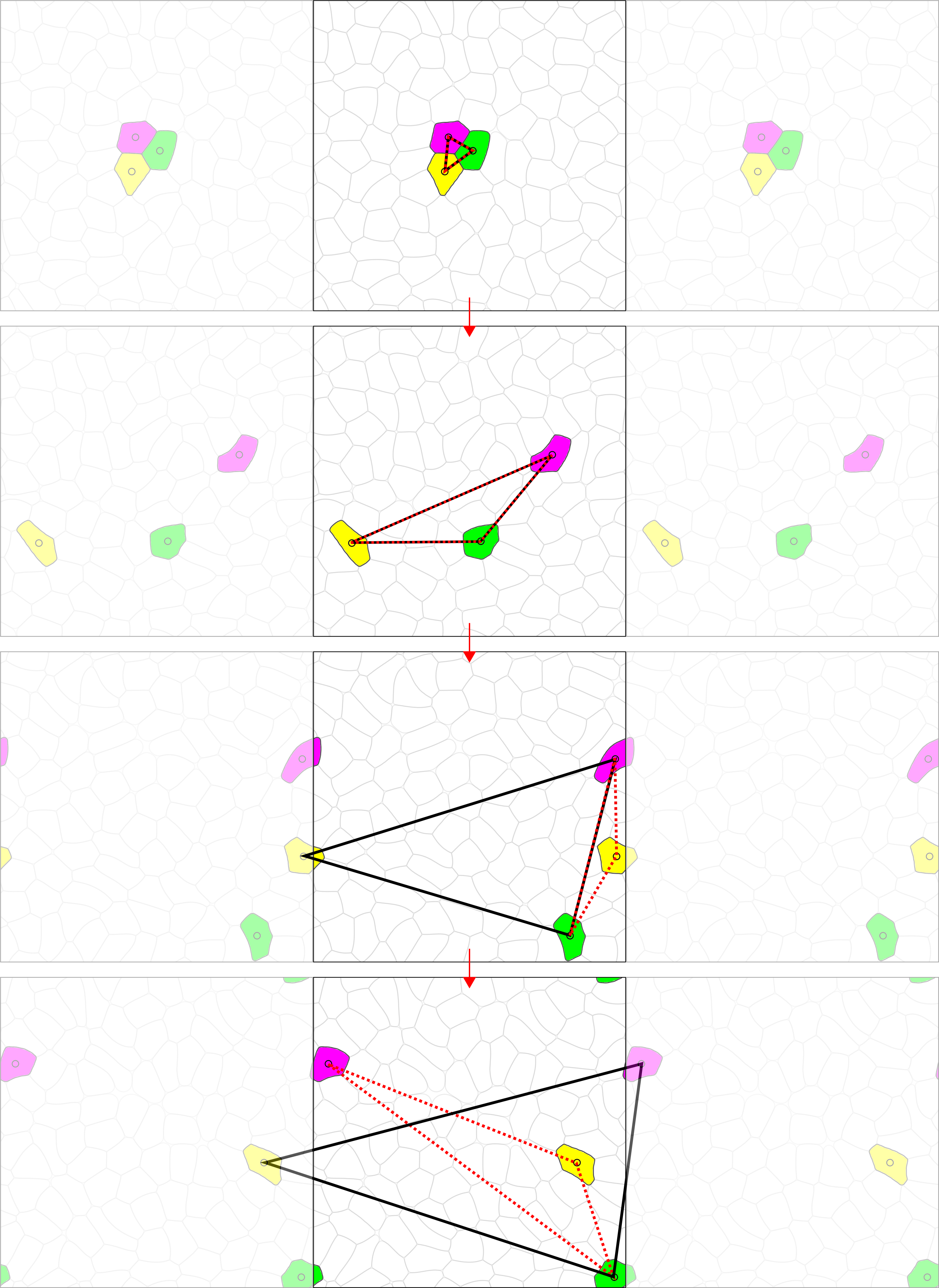}
    \caption{The triangle with red dotted sides is connected to the cells in the computational domain while, the triangle with black sides is connected to the location of the cells in an infinite 2D domain. Time evolves from top to bottom.}
    \label{fig:triangle_evolution_pbc}
\end{figure}
While the general notion of periodic boundary conditions is standard, the definition of the mean dispersion of a single cell ($\sigma^2$), cell pairs ($\Pi$) and triplets ($\Lambda$) requires some additional thoughts. We essentially assume that when a cell crosses the periodic boundary, it moves into an identical but a distinct copy of the domain that lies adjacent to the domain as illustrated in Figure \ref{fig:triangle_evolution_pbc}. The history of the cells motion is used to determine the modified location of center of mass in the resulting infinite domain.  

\begin{figure}
    \centering
    \includegraphics[width=.65\textwidth]{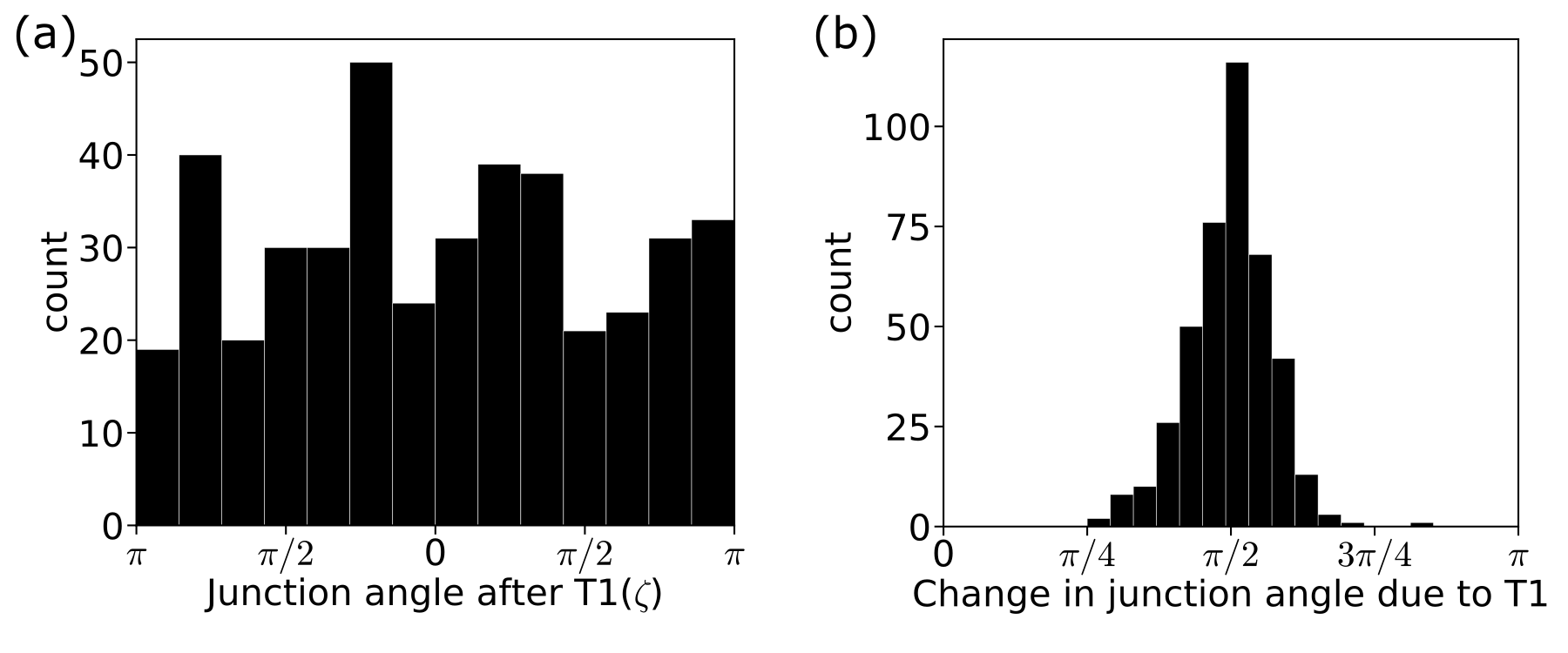}
    \caption{(a) Histogram of the junction angle after a T1 transition, the angle $\zeta$ is defined in Figure. 1 in the main text. (b) Histogram of change in junction angle before and after the T1 transition.}
    \label{fig:T1 orientation histogram}
\end{figure}
\section{Average flow field at T1 transition}
\begin{figure}
    \centering
    \includegraphics[width=.95\textwidth]{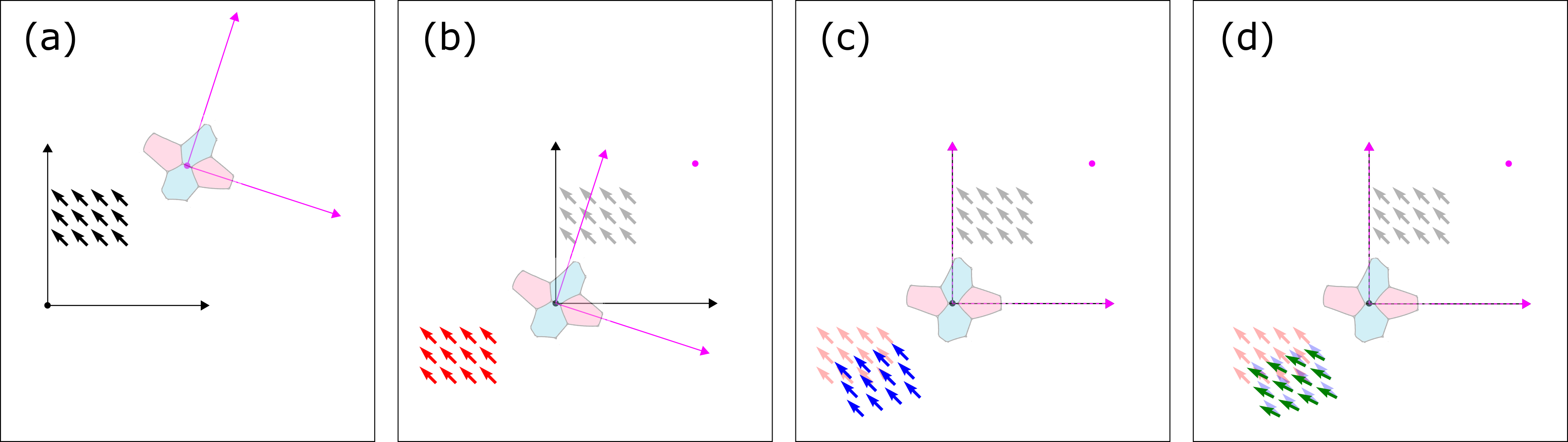}
    \caption{Steps involved in transformation of the velocity field to align it with respect to a T1 transition. (a) A small section of a field is shown in black. A T1 transition is shown with the cells that lose neighbours coloured in pink, while the cells that gain neighbours are coloured in blue. A magenta reference frame is attached to the epicenter of the T1 and is aligned with the orientation of the the T1, $\zeta$. (b) The origin is translated to the epicenter of the T1 transitions. (c) The frame attached to the origin is rotated to align the loser cells with the horizontal axis. (d) The orientation of the vector field is rotated by $\zeta_{T1}$}
    \label{fig:field transformations}
\end{figure}
\begin{figure}
    \centering
    \includegraphics[width=.70\textwidth]{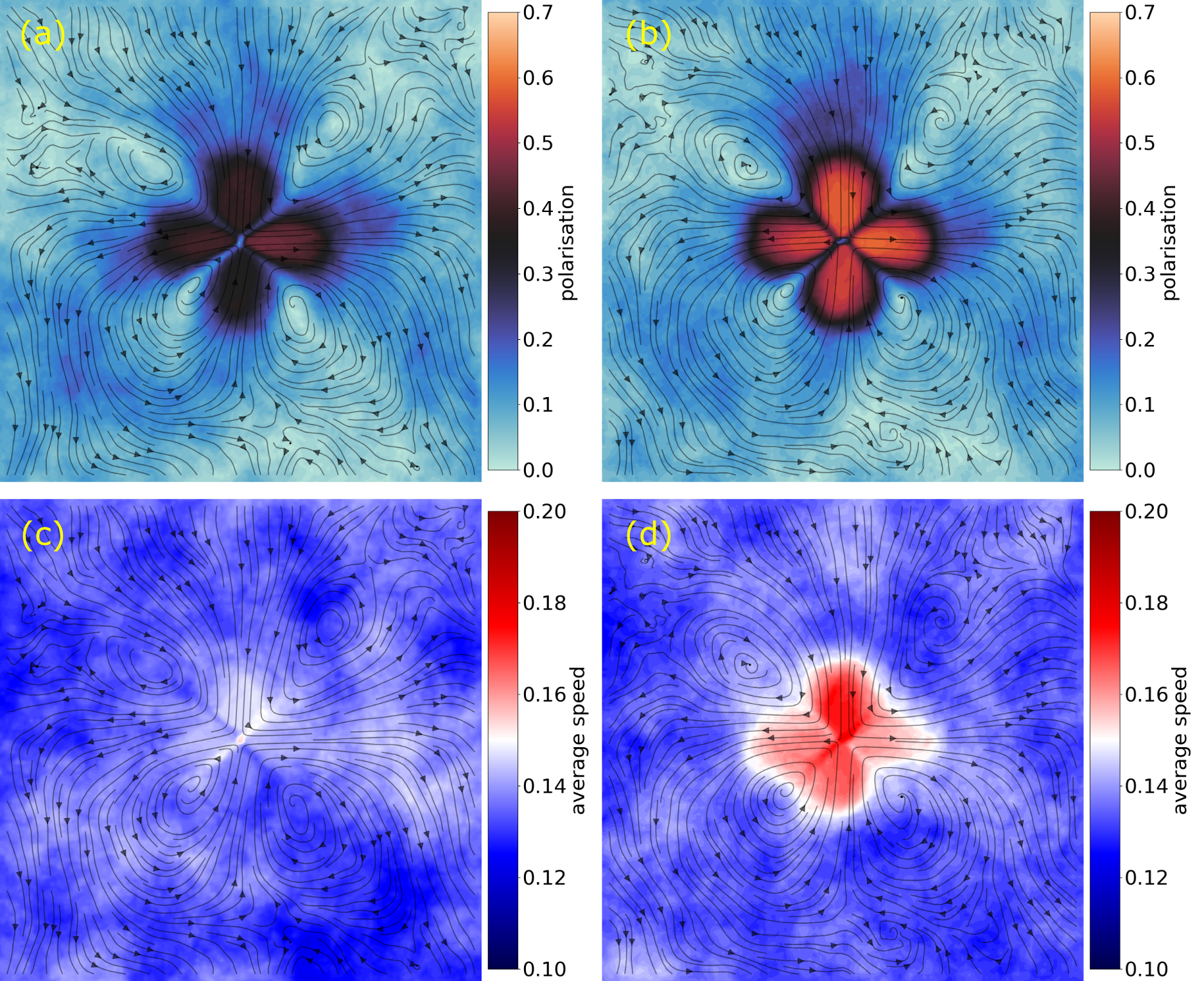}
    \caption{The streamlines in (a) and (c) corresponds to $\mathbf{v}_{T1}^{-}(\mathbf{x}, 0)$ whereas, the colormap corresponds to $\mathbf{p}_{T1}^{-}(\mathbf{x}, 0)$ and $v_{T1}^{-}(\mathbf{x}, 0)$, respectively. The streamlines in (b) and (d) corresponds to $\mathbf{v}_{T1}^{+}(\mathbf{x}, 0)$ whereas, the colormap corresponds to $\mathbf{p}_{T1}^{+}(\mathbf{x}, 0)$ and $v_{T1}^{+}(\mathbf{x}, 0)$, respectively. }
    \label{fig:flow fields before and after T1}
\end{figure}

Let, $N_{T1}$ be the total number of T1 transitions. Suppose, 4 cells, $A, B, C, D$,  undergo a T1 transition. Let the cell neighbour pairs involving the four cells before T1 be $(A,B), (A,C), (A,D), (C,B), (D, B)$ and after T1 be $(C,D), (A,C), (A,D), (C,B), (D, B)$. At the start of the T1 transition, cells $A$ and $B$ lose contact and at the end of the T1 transition, cells $C$ and $D$ make contact. We denote the start time and end time of the $m^{th}$ T1 transition as $t_{m, s}$ and $t_{m, e}$, respectively. Before $t_{m, s}$, the 3-way vertices where the cells $(A, B, C)$ and $(A, B, D)$ meet, move towards each other, shrinking the common junction between them, and eventually the junction vanishes at $t_{m,s}$. Between, $t_{m, s}$ and $t_{m, e}$, there is a extracellular gap between the four cells. After $t_{m,e}$, two new 3-way vertices where the cells $(A, C, D)$ and $(B, C, D)$ meet are created, and they move away from each other. Let, $\mathbf{s}_{m, 1}$ and $\mathbf{s}_{m, 2}$ be the positions of the two 3-way vertices a short time after $t_{m, e}$. We define the orientation of the $m^{th}$ T1 transition as $\zeta_{m} = \arg({\mathbf{s}_{m, 1} - \mathbf{s}_{m, 2}})$. Figure~\hyperref[fig:T1 orientation histogram]{S2a} shows the histogram for the orientation of T1 transitions. Figure~\hyperref[fig:T1 orientation histogram]{S2b} is the histogram of the difference in junction angle before and after the T1 transition. The junctions before and after a T1 transition are mostly perpendicular to each other.

For each T1 transition, we also find the location of its epicenter as in \cite{jainRobustStatisticalProperties2023a}. The epicenter is defined as the location of a point whose sum of distances to the four cells involved in the T1 transition is the minimum found at the time $\frac{t_{m, s}+t_{m, e}}{2}$. We denote the location of epicenter of $m^{th}$ T1 transition by $\mathbf{s}_{m, c}$. Essentially, it is the center of the gap formed between the four cells during the T1 transition. 

At any given time $t$, we define the global velocity field as 
\begin{equation}
\mathbf{v}_g(\mathbf{x}, t) = \sum_{i=1}^N \frac{(\phi_i+1)}{2}\frac{(\mathbf{r}_i(t) - \mathbf{r}_i(t'))}{t-t'}.
\end{equation}

For the $m^{th}$ T1 transition, at any time $t$, we can transform the global field such that 
\begin{equation}
    \mathbf{v}_{m}(\mathbf{x}, t) = \mathbf{R}_{\zeta_m}\mathbf{v}_{g}(\mathbf{R}_{\zeta_m}(\mathbf{x}-\mathbf{s}_{m, c}), t),
\end{equation}
where $\mathbf{R}_{\zeta_m}$ is the 2D rotation matrix with rotation angle $\zeta_m$. $\mathbf{v}_{m}(\mathbf{x}, t)$ is centered such that the epicenter is at the origin, and the junction formed after the T1 transition is aligned along the horizontal axis. Figure~\ref{fig:field transformations} visualises the  transformation from $\mathbf{v}_{g}(\mathbf{x}, t)$ to $\mathbf{v}_{m}(\mathbf{x}, t)$ in 3 steps for an arbitrary field. First, the field is translated such that the origin coincides with the epicenter of the T1 transition. Next, the grid is rotated such that the junction formed after the T1 transition aligns along the horizontal axis. Later, the vectors themselves are rotated while being fixed in space to reflect the right orientation with respect to the epicenter. The mean velocity field around a T1 transition at time $\tau$ after the end of a T1 transition is then given as 
\begin{equation}
    \mathbf{v}_{T1}^{+}(\mathbf{x}, \tau)  = \frac{1}{N_{T1}}\sum_{m=1}^{N_{T1}} \mathbf{v}_{m}(\mathbf{x}, t_{m, e}+\tau). 
\end{equation}
The corresponding mean velocity magnitude field is obtained as 
\begin{equation}
    v_{T1}^{+}(\mathbf{x}, \tau)  = \frac{1}{N_{T1}}\sum_{m=1}^{N_{T1}} |\mathbf{v}_{m}(\mathbf{x}, t_{m, e}+\tau)|.
\end{equation}
The corresponding polarisation field around a T1 transition is defined as 
\begin{equation}
    \mathbf{p}_{T1}^{+}(\mathbf{x}, \tau)  = \frac{1}{N_{T1}}\sum_{m=1}^{N_{T1}} \frac{\mathbf{v}_{m}(\mathbf{x},  t_{m, e}+\tau)}{|\mathbf{v}_{m}(\mathbf{x},  t_{m, e}+\tau)|}. 
\end{equation}
Likewise, the fields before the start of a T1 transition are defined as follows
\begin{eqnarray}
    \mathbf{v}_{T1}^{-}(\mathbf{x}, \tau)  &=& \frac{1}{N_{T1}}\sum_{m=1}^{N_{T1}} \mathbf{v}_{m}(\mathbf{x}, t_{m, s}-\tau), \\
    v_{T1}^{-}(\mathbf{x}, \tau)  &=& \frac{1}{N_{T1}}\sum_{m=1}^{N_{T1}} |\mathbf{v}_{m}(\mathbf{x}, t_{m, s}-\tau)|
\end{eqnarray}
and 
\begin{equation}
    \mathbf{p}_{T1}^{-}(\mathbf{x}, \tau)  = \frac{1}{N_{T1}}\sum_{m=1}^{N_{T1}} \frac{\mathbf{v}_{m}(\mathbf{x},  t_{m, s}-\tau)}{|\mathbf{v}_{m}(\mathbf{x},  t_{m, s}-\tau)|}. 
\end{equation}
The streamlines in the Figure~\hyperref[fig:flow fields before and after T1]{S4a} and Figure~\hyperref[fig:flow fields before and after T1]{S4c} corresponds to $\mathbf{v}_{T1}^{-}(\mathbf{x}, 0)$ whereas, the colormap corresponds to $\mathbf{p}_{T1}^{-}(\mathbf{x}, 0)$ and $v_{T1}^{-}(\mathbf{x}, 0)$, respectively. Likewise, the streamlines in the Figure~\hyperref[fig:flow fields before and after T1]{S4b} and Figure~\hyperref[fig:flow fields before and after T1]{S4d}(also Figures 1d and 1e in the main text) corresponds to $\mathbf{v}_{T1}^{+}(\mathbf{x}, 0)$ whereas, the colormap corresponds to $\mathbf{p}_{T1}^{+}(\mathbf{x}, 0)$ and $v_{T1}^{+}(\mathbf{x}, 0)$, respectively. 
The supplementary movies 1 and 3 show $\mathbf{p}_{T1}^{-}(\mathbf{x}, \tau)$ and $v_{T1}^{-}(\mathbf{x}, \tau)$, respectively, and the supplementary movies 2 and 4 show $\mathbf{p}_{T1}^{+}(\mathbf{x}, \tau)$ and $v_{T1}^{+}(\mathbf{x}, \tau)$, respectively.

\section{Multiple T1 transitions are required for triplet area to diverge}
\begin{figure}
    \centering
    \includegraphics[width=.50\textwidth]{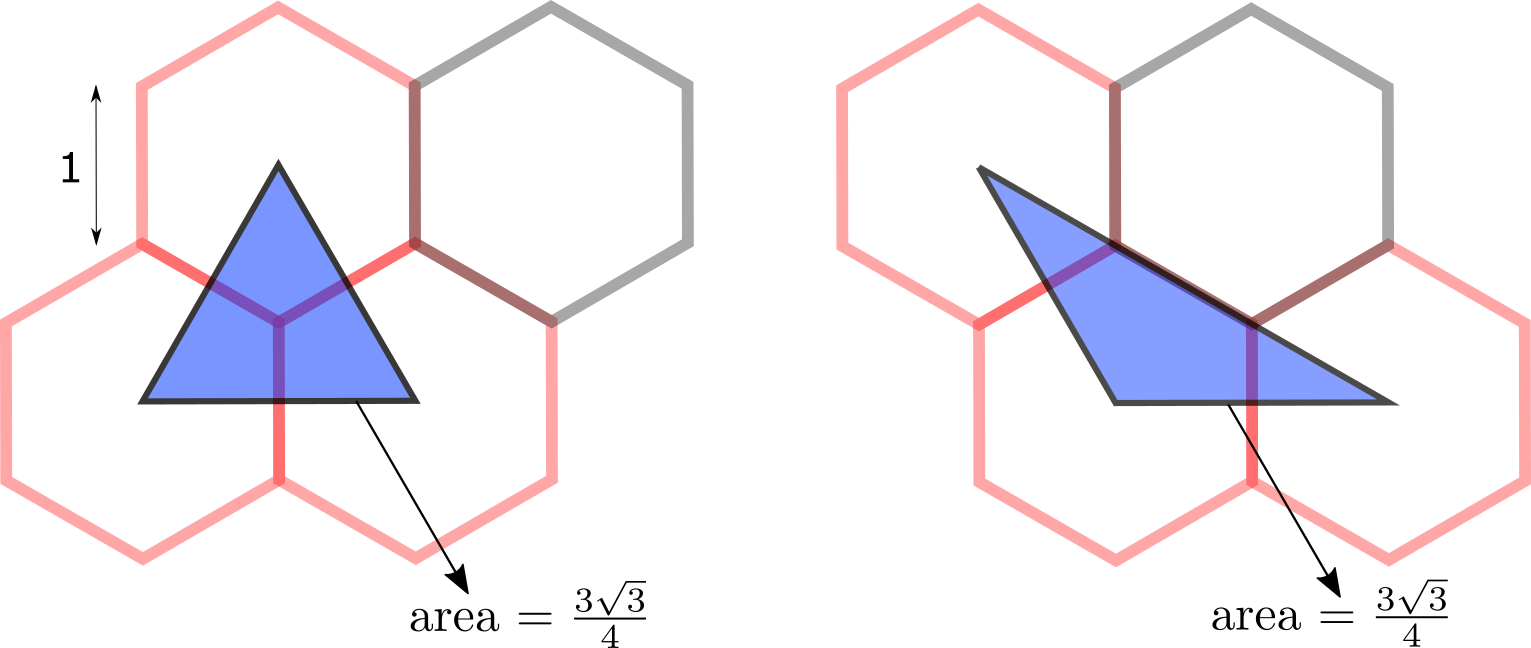}
    \caption{A triangle joining $3$ red cells is shown when (left) all $3$ cells are mutually adjacent and when (right) one pair are not neighbours.}
    \label{fig:first T1 triangle area}
\end{figure}

Figure~\ref{fig:first T1 triangle area} shows an illustration of a section of hexagonal lattice comprising of hexagonal cells. A triangle is marked that joins the cell centers of the $3$ red cells. In the left figure, all $3$ cells are mutually adjacent, while on the right figure, $2$ of the $3$ cells are not in contact. The area of the triangles are the same in both situations. For a triplet area to diverge significantly, at least one of the $3$ cells must break contact from the other $2$ cells. This is why, multiple T1 transitions are required before the mean triplet-separation area ($\lambda$) starts to increase.

\section{Two modes of collective behaviour}
\begin{figure}[]
    \includegraphics[width=0.65\textwidth]{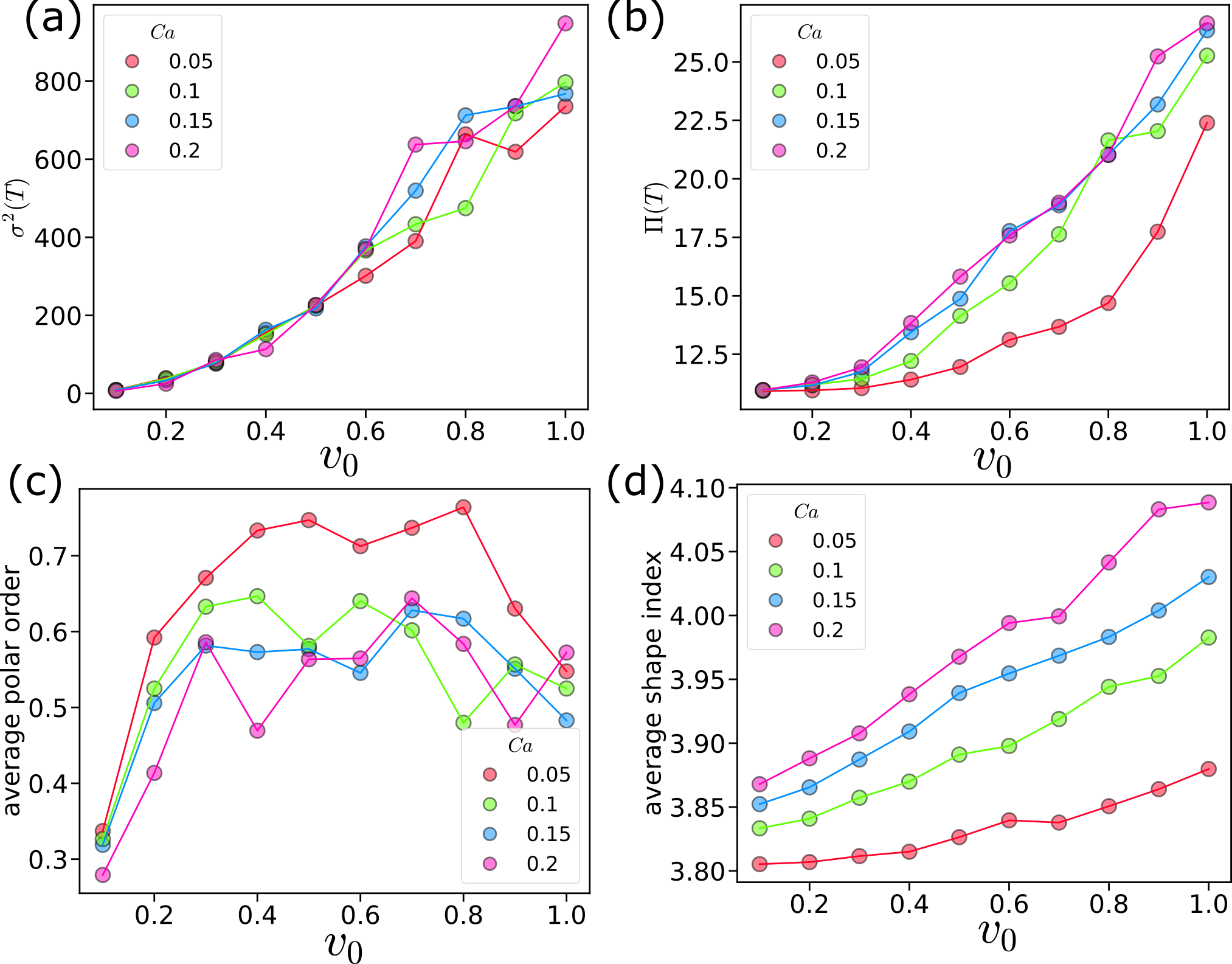}
    \caption{ Various properties are plotted while varying activity $v_0$ and deformability $Ca$. (a) Mean squared displacement and (b) Mean pair separation for the time lag $T=150$. (c) average polar order. (d) average shape index.}
    \label{fig:Figure 6}
\end{figure}

Figures~\hyperref[fig:Figure 6]{S6a} and ~\hyperref[fig:Figure 6]{S6b} show the self dispersion ($\sigma(T)$) and relative dispersion properties ($\Pi(T)$). With increase in activity, $\sigma(T)$ increases. However, there is no change with $Ca$. This suggests, that the distance travelled by cells irrespective of their deformability is similar. Whereas for stiffer cells i.e. low $Ca$, $\Pi(T)$ is lower suggesting that stiffer cells move less relative to each other. 

The polar order at any given time is found as $$\psi(t) = \frac{1}{N}\left|\sum_{i=1}^{N} \frac{(\mathbf{r}_i(t) - \mathbf{r}_i(t'))}{t-t'}\right|,$$ and is a measure of collective flocking behaviour. If $\psi(t)=1$, all cells move in the same direction, while if $\psi(t)=0$, the cell velocities have random orientation. The average polar order is the temporal average of $\psi(t)$. We see that stiffer cells have higher average polar order, see Figure~\hyperref[fig:Figure 6]{S6c}. This explains the discrepancy where we see stiff cells move similar distances as soft cells as they have a slightly higher flocking tendency. To further quantify the differences due to varying $Ca$, we plot the average shape index in Figure~\hyperref[fig:Figure 6]{S6d}. Shape index is the ratio of the perimeter of cells to the square root of area of the cells. A circle has the lowest shape index, and elongated cells have higher shape index.

\bibliography{mainT1flow2}